\documentclass[onecolumn,aps,prc,showpacs,superscriptaddress]{revtex4}
\usepackage{psfig}

\newcommand {\pt}	{p_{T}}
\newcommand {\phit}	{\phi_{t}}
\newcommand {\dphi}	{\Delta\phi}
\newcommand {\cij}	{\cos2\dphi_{ij}}
\newcommand {\cphi}	{\cos2\dphi}
\newcommand {\tphi}	{\tilde\phi}
\newcommand {\intpi}	{\int_{0}^{2\pi}}
\newcommand {\vtwo}[1]	{v_{2,#1}}
\newcommand {\vv}[1]	{v_2\{\rm #1\}}
\newcommand {\vsq}[1]	{v_2^2\{#1\}}
\newcommand {\mean}[1]	{\langle#1\rangle}
\newcommand {\acl}	{{d}}
\newcommand {\ajet}	{{d,jet}}
\newcommand {\Nt}	{N_{t}}
\newcommand {\ns}	{{ns}}
\newcommand {\as}	{{as}}
\newcommand {\Nhy}	{N_{hy}}
\newcommand {\Nbg}	{N_{bg}}
\newcommand {\Ncl}	{N_{cl}}
\newcommand {\Na}	{N_{\acl}}
\newcommand {\Njet}	{N_{\ajet}}
\newcommand {\Nak}	{N_{d,k}}
\newcommand {\fcl}	{f_{\acl}}
\newcommand {\rhocl}	{\rho_{cl}}
\newcommand {\be}	{\begin{equation}}
\newcommand {\ee}	{\end{equation}}
\newcommand {\bea}	{\begin{eqnarray}}
\newcommand {\eea}	{\end{eqnarray}}

\begin{document}

\title{Identification of flow-background to subtract in jet-like azimuthal correlation}
\author{Quan Wang}
\author{Fuqiang Wang}
\affiliation{Department of Physics, Purdue University, 525 Northwestern Ave., West Lafayette, IN 47907}

\begin{abstract}
We derive an analytical form for flow-background to jet-like azimuthal correlation in a cluster approach. We argue that the elliptic flow parameter to use in jet-correlation background is that from two-particle method excluding non-flow correlation unrelated to the reaction plane, but including cross-terms between cluster correlation and cluster flow. We verify our result with Monte Carlo simulations. We discuss implications of our finding in the context of jet-like correlations from STAR and PHENIX.
\end{abstract}


\maketitle

\section{Introduction}

Jet-like angular correlation studies with high transverse momentum ($\pt$) trigger particles have provided valuable information on the properties of the medium created in relativistic heavy-ion collisions~\cite{STARwp,PHENIXwp}. In such studies, correlation functions are formed in azimuthal angle difference between an associated particle and a high $\pt$ trigger particle, which preferentially selects (di-)jet. One important aspect of these studies is the subtraction of combinatorial background which itself is non-uniform due to anisotropic particle distribution with respect to the reaction plane-- both the trigger particle and the associated particles are correlated with the common reaction plane in an event. One critical part is to determine flow parameters, mainly elliptic flow ($v_2$), to use in constructing background. 

There are many $v_2$ measurements~\cite{v4,v2data}. They contain various degrees of non-flow contributions, such as those from resonance decays and jet correlations. 
Those non-flow effects should not be included in the background to subtract from jet-like correlations. 
We shall refer to this jet-correlation background as flow-background. 
The anisotropic flow to be used for flow-background should be ideally that from two-particle method, $\vv{2}$~\cite{v2method,v2method2}, because jet-like correlation is analyzed by two-particle correlation method. Moreover, two-particle anisotropic flow contains fluctuations which should be included in jet-correlation flow-background~\cite{v2method,v2method2}. 

Non-flow is due to azimuthal correlations unrelated to the reaction plane, such as resonances, (mini)jets, or generally, clusters. Non-flow in two-particle $\vv{2}$ was studied in a cluster approach~\cite{FWang}; analytical form was derived for each non-flow component. In this paper, we shall demonstrate that the flow to be used in jet-correlation background subtraction should be the two-particle $\vv{2}$ excluding cluster correlations unrelated to the reaction plane, but including cross-terms between cluster correlations and cluster flow. We verify our result with Monte Carlo simulations. We then examine available jet-like correlation data from STAR~\cite{STARwp} and PHENIX~\cite{PHENIXwp} with our refined flow-background. Finally we discuss implications of our result on the observed conical emission signal~\cite{3part} and on medium modification of jets in general.

\section{Elliptic flow for jet-correlation background}

In this section, we derive an analytical form for flow-background to jet-correlation in the cluster approach, as used in our non-flow study~\cite{FWang}. We suppose a relativistic heavy-ion collision event is composed of hydrodynamic medium particles, jet-correlated particles, and particles correlated via clusters. Hydro-particles, high $\pt$ trigger particles, and clusters are distributed relative to reaction plane ($\psi$) by
\be
\frac{dN}{d\phi}=\frac{N}{2\pi}\left[1+2v_2\cos2(\phi-\psi)\right]
\label{eqn1}
\ee	 
with the corresponding elliptic flow parameter $v_2$ and multiplicity $N$. Particle azimuthal distribution with respect to a trigger particle is
\be
\frac{1}{\Nt}\frac{dN}{d\dphi}=\frac{d\Nhy}{d\dphi}+\sum_{k\neq jet\in clus}\frac{d\Nak}{d\dphi}+\sum_{k\in jet}\frac{d\Nak}{d\dphi}+\frac{d\Njet}{d\dphi}
\label{eqn2}
\ee	 
where $\dphi=\phi-\phit$. In Eq.~(\ref{eqn2}), $d\Njet/d\dphi$ is jet-correlation signal of interest. All other terms are backgrounds; $\Nak$ is the number of daughter particles in cluster $k$. 

If trigger particle multiplicity is Poisson and effects due to interplay between collision centrality selection (usually via multiplicity) and trigger bias are negligible, then the background event of the triggered (di-)jet should be identical to any inclusive event, without requiring a high $\pt$ trigger particle, but with all other event selection requirements as for the triggered event~\cite{3part_method}. Therefore, experimentally one can use inclusive events 
to obtain flow-background:
\be
\frac{1}{\Nt}\frac{dN}{d\dphi}=a\left(\frac{d\Nhy}{d\dphi}+\sum_{k\neq jet\in clus}\frac{d\Nak}{d\dphi}+\sum_{k\in jet}\frac{d\Nak}{d\dphi}\right)_{inc}+\frac{d\Njet}{d\dphi}
\ee
where $a$ is a normalization factor, often determined by the assumption of ZYAM or ZYA1 (zero jet-correlated yield at minimum or at $\dphi=1$)~\cite{zyam,jetspec}, and is approximately unity. 

In this paper we are interested in the anisotropic flow to be used in jet-correlation background. So we will not concern ourselves with the background normalization, but only the background shape. We rewrite the background in Eq.~(\ref{eqn2}) as
\be
\frac{d\Nbg}{d\dphi}=\frac{d\Nhy}{d\dphi}+\sum_{k\neq jet\in clus}\frac{d\Nak}{d\dphi}+\sum_{k\in jet}\frac{d\Nak}{d\dphi}=\frac{d\Nhy}{d\dphi}+\sum_{cl}\Ncl\frac{d\Na}{d\dphi}
\ee
where we have eliminated subscript `$inc$' to lighten notation. We have summed over all cluster types `$cl$' including jet-correlation, where $\Ncl$ is the number of clusters of type `$cl$'. Different cluster types include jet and minijet correlations, resonance decays, etc.

The hydro-background is simply
\be
\frac{d\Nhy}{d\dphi}=\frac{\Nhy}{2\pi}\left(1+2\vtwo{t}\vtwo{hy}\cphi\right)
\label{eqn5}
\ee
where $\vtwo{t}$ is elliptic flow parameter of trigger particles and $\vtwo{hy}$  is that of hydro-medium particles. 

The cluster particles background is given by
\be
\frac{d\Na}{d\dphi}=\intpi d\tphi_t\rho_t(\tphi_t)\intpi d\tphi_k\rhocl(\tphi_k)\intpi d\dphi_i\fcl(\dphi_i,\tphi_k)\times\frac{1}{2\pi}\delta(\dphi_i+\tphi_k-\dphi-\tphi_t)
\label{eqn6}
\ee
where $\tphi_t=\phi_t-\psi$, $\tphi_k=\phi_k-\psi$, $\dphi_i=\phi_i-\phi_k$, and $\rho_t(\tphi_t)=\frac{1}{2\pi}\left(1+2v_{2,t}\cos2\tphi_t\right)$ and $\rhocl(\tphi_k)=\frac{1}{2\pi}\left(1+2v_{2,cl}\cos2\tphi_k\right)$ are density profiles (i.e., $v_2$-modulated distributions) of trigger particles and clusters relative to the reaction plane, respectively. We have assumed that the cluster axis (or cluster parent) distribution is also anisotropic with respect to the reaction plane. In Eq.~(\ref{eqn6}), $\fcl(\dphi_i,\tphi_k)=\frac{d\Nak}{d\dphi_i}\equiv\frac{d\Na(\tphi_k)}{d\dphi_i}$ is distribution of daughter particles in cluster relative to cluster axis (cluster correlation function), which may depend on the cluster axis relative to the reaction plane $\tphi_k$~\cite{Aoqi}. Decomposing $\rho_t(\tphi_t)$, we obtain
\be
\frac{d\Na}{d\dphi}=\frac{1}{2\pi}\intpi d\tphi_k\rhocl(\tphi_k)\intpi d\dphi_i\fcl(d\dphi_i,\tphi_k)+\frac{2\vtwo{t}}{2\pi}\intpi d\tphi_k\rhocl(\tphi_k)\intpi d\dphi_i\fcl(\dphi_i,\tphi_k)\cos2(\dphi_i+\tphi_k-\dphi).
\ee
Because of symmetry, $\fcl(\dphi_i,\tphi_k)=\fcl(-\dphi_i,-\tphi_k)$ and $\rhocl(\tphi_k)=\rhocl(-\tphi_k)$, we have $\intpi d\tphi_k\rhocl(\tphi_k)\intpi d\dphi_i\fcl(\dphi_i,\tphi_k)\sin2(\dphi_i+\tphi_k)=0$. Therefore
\be
\frac{d\Na}{d\dphi}=\frac{1}{2\pi}\intpi d\tphi_k\rhocl(\tphi_k)\Na(\tphi_k)+\frac{2\vtwo{t}}{2\pi}\cphi\intpi d\tphi_k\rhocl(\tphi_k)\intpi d\dphi_i\fcl(\dphi_i,\tphi_k)\cos2(\dphi_i+\tphi_k).
\label{eqn6p5}
\ee
Realizing that elliptic flow parameter of particles from clusters is given by
\be
\vtwo{\acl}\equiv\mean{\cos2(\phi-\psi)}_{cl}=\frac{1}{\Na}\intpi d\tphi_k\rhocl(\tphi_k)\intpi d\dphi_i\Na(\tphi_k)\fcl(\dphi_i,\tphi_k)\cos2(\dphi_i+\tphi_k).
\label{eqn6p7}
\ee
we rewrite Eq.~(\ref{eqn6p5}) into
\be
\frac{d\Na}{d\dphi}=\frac{\Na}{2\pi}\left(1+2\vtwo{t}\vtwo{\acl}\cphi\right).
\label{eqn7}
\ee

From Eq.~(\ref{eqn5}) and (\ref{eqn7}) we obtain the total background as given by
\be
\frac{d\Nbg}{d\dphi}=\frac{\Nbg}{2\pi}\left[1+2\vtwo{t}\left(\frac{\Nhy}{\Nbg}\vtwo{hy}+\sum_{cl}\frac{\Ncl\Na}{\Nbg}\vtwo{\acl}\right)\cphi\right],
\label{eqn8}
\ee
where 
\be
\Nbg=\Nhy+\sum_{cl}\Ncl\Na.
\ee
The $v_2$'s in Eqs.~(\ref{eqn5}), (\ref{eqn7}), and (\ref{eqn8}) include fluctuations, so they should be replaced by $\sqrt{\mean{v_2^2}}$. The hydro-particles $\sqrt{\mean{v_2^2}}$ is equivalent to two-particle $\vv{2}$ because there is no non-flow effect between hydro-particle pairs; same for the cluster $\sqrt{\mean{v_2^2}}$ because there is no non-flow effect between different clusters (we consider sub-clusters to be part of their parent cluster). Thus Eq.~(\ref{eqn8}) should be
\be
\frac{\Nbg}{d\dphi}=\frac{\Nbg}{2\pi}\left(1+2\vtwo{t}\vtwo{bg}\cphi\right)
\ee
where
\be
\vtwo{bg}=\frac{\Nhy}{\Nbg}\vv{2}_{hy}+\sum_{cl}\frac{\Ncl\Na}{\Nbg}\vv{2}_{\acl}.
\label{eqn11}
\ee
We note that here cluster includes single-particle (within a give $\pt$ range) cluster, which generally is part of a parent cluster including particles of all $\pt$. Those single-particle clusters do not contribute to non-flow in $\vv{2}_{\acl}$, but they differ from single hydro-particles because they may possess different $v_2$ values.

In principle, $\vtwo{t}$ should have a similar expression as Eq.~(\ref{eqn11}) out of symmetry reason:
\be
\vtwo{t}=\frac{N_{t,hy}}{N_{t,tot}}\vv{2}_{t,hy}+\sum_{cl\_t}\frac{N_{cl\_t}N_{t,cl\_t}}{N_{t,tot}}\vv{2}_{t,cl\_t}.
\ee
where $N_{t,hy}$ is number of high $\pt$ trigger particles from hydro-medium (i.e., background trigger particles), $\vv{2}_{t,hy}$ is the elliptic anisotropy of those background trigger particles, $N_{cl\_t}$ is number of clusters of type `$cl\_t$' containing at least one trigger particle, $N_{t,cl\_t}$ is number of trigger particles per cluster, $\vv{2}_{t,cl\_t}$ is elliptic flow parameter of trigger particles from clusters, and $N_{t,tot}=N_{t,hy}+\displaystyle{\sum_{cl\_t}}N_{cl\_t}N_{t,cl\_t}$. The only difference is that trigger particles are dominated by clusters (mostly jets), and those clusters are dominated by single-trigger-particle clusters; hydro-medium contribution to trigger particle population should be small. We note that jet-correlation functions are usually normalized by total number of trigger particles including those from hydro-medium background.

If particle correlation in clusters does not vary with cluster axis relative to the reaction plane,
\be
\vv{2}_{\acl}\equiv\vv{2}_{cl}\mean{\cphi}_{cl},
\label{eqn12}
\ee
and
\be 
\vtwo{bg}=\frac{\Nhy}{\Nbg}\vv{2}_{hy}+\sum_{cl}\frac{\Ncl\Na}{\Nbg}\vv{2}_{cl}\mean{\cphi}_{cl}.
\label{eqn13}
\ee

\section{Two-particle $v_2$ in cluster model}

Obviously, the elliptic flow in Eq.~(\ref{eqn11}) or (\ref{eqn13}) contains not only the two-particle anisotropy relative to the reaction plane, but also non-flow related to angular spread of clusters. How to obtain the elliptic flow as in Eq.~(\ref{eqn11}) or (\ref{eqn13})? In~\cite{FWang} we have derived two-particle $\vv{2}$ in a general hydro+cluster approach:
\be
\vsq{2}=\left(\frac{\Nhy}{\Nbg}\vv{2}_{hy}+\sum_{cl}\frac{\Ncl\Na}{\Nbg}\vv{2}_{\acl}\right)^2+\sum_{cl}\frac{\Ncl\Na^2}{\Nbg^2}\left(\mean{\cij}_{cl}-\vsq{2}_{\acl}\right).
\label{eqn14}
\ee

The quantity in the first pair of parentheses in r.h.s.~of Eq.~(\ref{eqn14}) is elliptic flow due to correlation with respect to the reaction plane. The second term in the r.h.s.~arises from cluster correlation~\cite{FWang}; the small correction $\vsq{2}_{\acl}$ is due to assumptions of Poisson statistics for number of clusters and particle multiplicity in clusters, but not the product of the two~\cite{FWang}. Since elliptic flow is formally defined to be relative to the reaction plane, the first term in r.h.s.~of Eq.~(\ref{eqn14}) may be considered as ``true'' elliptic flow (except flow fluctuation effect), $\vtwo{{\rm flow}}$. We note, however, it is not necessarily as same as hydro-flow because of contamination from clusters due to coupling between cluster correlation and cluster flow. The second term in r.h.s.~of Eq.~(\ref{eqn14}) can be considered as non-flow, $\vtwo{{\rm non-flow}}$; non-flow is due to correlations between particles from the same dijet or the same cluster. Eq.~(\ref{eqn14}) can be expressed into
\be
\vsq{2}=v^2_{2,{\rm flow}}+v^2_{2,{\rm non-flow}}.
\ee

Comparing Eq.~(\ref{eqn14}) with Eq.~(\ref{eqn11}), we see that 
\be
\vtwo{bg}=\vtwo{{\rm flow}},
\ee
i.e., the quantity in the first pair of parentheses in r.h.s.~of Eq.~(\ref{eqn14}) is the $v_2$ parameter in Eq.~(\ref{eqn11}) that is needed in constructing jet-correlation background. In other words, elliptic flow parameter that should be used in jet-correlation flow background is the ``true'' two-particle elliptic flow (i.e., due to the reaction plane and including fluctuation).

\section{Monte Carlo checks}

In this section, we verify our analytical result by Monte Carlo simulations. We generate events consisting of three components. One component is hydro-medium particles according to Eq.~(\ref{eqn1}), given hydro-particles elliptic flow parameter $\vtwo{hy}$ and Poisson distributed number of hydro-particles with average multiplicity $\Nhy$. The second component is clusters, given cluster elliptic flow parameter $\vtwo{cl}$ and Poisson distributed number of clusters with average $\Ncl$; each cluster is made of particles with Poisson multiplicity distribution with average $\Na$ and Gaussian azimuth spread around cluster axis with $\sigma_{\acl}$. The third component is trigger particles with accompanying associated particles; the trigger particle multiplicity is Poisson with average $\Nt$, and the elliptic flow parameter is $\vtwo{t}$. The associated particles are generated for each trigger particle by correlation function:
\bea
f(\dphi,\tphi_t)&=&C(\tphi_t)+\frac{N_{\ns}(\tphi_t)}{\sqrt{2\pi}\sigma_{\ns}(\tphi_t)}\exp\left[-\frac{(\dphi)^2}{2\sigma_{\ns}^2(\tphi_t)}\right]+\nonumber\\
&&\frac{N_{\as}(\tphi_t)}{\sqrt{2\pi}\sigma_{\as}(\tphi_t)}\left(\exp\left[-\frac{\left(\dphi-\pi+\theta(\tphi_t)\right)^2}{2\sigma_{\as}^2(\tphi_t)}\right]+\exp\left[-\frac{\left(\dphi-\pi-\theta(\tphi_t)\right)^2}{2\sigma_{\as}^2(\tphi_t)}\right]\right),
\label{eqn16}
\eea
where the near- and away-side associated particle multiplicities are Poisson with averages $N_{\ns}(\tphi_t)$ and $N_{\as}(\tphi_t)$, respectively. The Gaussian widths of the near- and away-side peaks are fixed, and the two away-side symmetric peaks are set equal and their separation is fixed. All parameters in the jet-correlation function of Eq.~(\ref{eqn16}) can be dependent on the trigger particle azimuth relative to the reaction plane, $\tphi_t$.

We first verify Eq.~(\ref{eqn14}) by generating events with hydro-particles and jet-correlated particles. (We do not include other clusters except jet-correlations.) We use $\Nhy=150$, $\vtwo{hy}=0.05$, and $\Nt=2$, $\vtwo{t}=0.5$. We use the large trigger particle $v_2$ in order to maximize the effect of non-flow. For jet-correlation function, we generate back-to-back dijet with $N_{\ns}=0.7$, $N_{\as}=1.2$, $\sigma_{\ns}=0.4$, $\sigma_{\as}=0.7$, and $\theta=0$ (referred to as dijet model). We fix $v_2$ in the simulation, i.e., $v_2$ fluctuation is not included. We simulate $10^6$ events and calculate $\vsq{2}=\mean{\cij}$. Including only hydro-particles, we obtain $\vv{2}_{hy}=0.05005\pm0.00009$, consistent with the input. Including all simulated events and all particles (hydro-particles and jet-correlated particles), we obtain $\vv{2}_{inc}=0.05560\pm0.00008$. Using triggered events (events containing at least one trigger particle) only, we obtain $\vv{2}_{trig\_evt}=0.05642\pm0.00008$. Using triggered events but excluding one dijet at a time (i.e., using the underlying background event of the dijet) and repeating over all dijets in the event, we obtain $\vv{2}_{bg}=0.05568\pm0.00008$. We see that the background $v_2$ is as same as that obtained from inclusive events, $\vv{2}_{bg}=\vv{2}_{inc}$, and both are smaller than that from triggered events only.

We can in fact predict the inclusive event $v_2$ by Eq.~(\ref{eqn13}) using the ``hydro + dijet'' model. The average $\sqrt{\mean{\cij}}$ of jet-correlated particle pairs within the same dijet is $\sqrt{\mean{\cij}}_{jet}=\mean{\cphi}_{jet}=0.5054\pm0.0004$. This is consistent with the expected value 
\[\mean{\cphi}_{jet}=\frac{N_{\ns}}{N_{\ns}+N_{\as}}\exp\left(-2\sigma_{\ns}^2\right)+\frac{N_{\as}}{N_{\ns}+N_{\as}}\exp\left(-2\sigma_{\as}^2\right)\cos2\theta=0.5046\]
where $\theta=0$.
The average $\sqrt{\mean{\cij}}$ of pairs of particles from different dijets is $0.2516\pm0.0008$; it equals to $\vv{2}_{\ajet}=\vtwo{t}\mean{\cphi}_{jet}=0.5\times0.5046=0.2523$. The average $\sqrt{\mean{\cij}}$ for cross-talk pairs of background particle and jet-correlated particle is $0.10064\pm0.00004$; and it equals to the expected value $\sqrt{\vv{2}_{hy}\vtwo{t}\mean{\cphi}_{jet}}=\sqrt{0.05\times0.5\times0.5046}=0.1123$. The inclusive event two-particle elliptic flow parameter is $\vv{2}=\sqrt{\left(\frac{150}{153.8}\times0.05+\frac{3.8}{153.8}\times0.2523\right)^2+\frac{2\times1.9^2}{153.8^2}\left(0.5054^2-0.2523^2\right)}=
0.05553$; this is indeed consistent with $\vv{2}_{inc}$ or $\vv{2}_{bg}$ obtained from simulation.

We now verify Eq.~(\ref{eqn11}) or (\ref{eqn13}) as the correct $v_2$ to be used for jet-correlation background subtraction. We generate Poisson distributed hydro-particles with average multiplicity $\Nhy=150$ and fixed elliptic flow parameter $\vtwo{hy}=0.05$. We generate Poisson distributed trigger particles with average trigger particle multiplicity $\Nt=2.0$; we use different jet-correlation functions (discussed below). We also include clusters that do not have trigger particles (referred to as minijet clusters); the particle multiplicity per minijet cluster is Poisson distributed with average $\Na=5$, and the number of minijet clusters is also Poisson distributed but we vary the average number of clusters $\Ncl$; we fix the cluster shape to be Gaussian with width $\sigma_{\acl}=0.5$ (and average angular spread $\mean{\cphi}_{cl}=\exp\left(-2\sigma^2_{\acl}\right)=0.6065)$, and also fix the cluster elliptic flow parameter $\vtwo{cl}=0.20$. We simulate $10^6$ events and form raw correlation functions normalized by the number of trigger particles. In order to extract the real background $v_2$ from the simulations, we subtract the input jet-correlation function. If the jet-correlation function varies with the trigger particle angle relative to the reaction plane, the trigger multiplicity weighted average jet-correlation function is subtracted. We fit the resultant background function to $B\left(1+2\vtwo{t}\vtwo{{\rm fit}}\cphi\right)$ where $B$ and $\vtwo{{\rm fit}}$ are fit parameters. We treat the input $\vtwo{t}$ as known; we did not include any complication into $\vtwo{t}$. We compare the fit $\vtwo{{\rm fit}}$ to the calculated one by Eq.~(\ref{eqn11}) or (\ref{eqn13}). We study several cases with different shapes for jet-correlation function, as well as varying values for some of the input parameters: 
\begin{itemize}
\item[(i)] ``hydro + dijet'' model: we generate back-to-back dijets accompanying trigger particles, without other clusters. The calculated $\vtwo{bg}$ by Eq.~(\ref{eqn13}) is $\vtwo{bg}=\frac{150}{153.8}\times0.05+\frac{2\times1.9}{153.8}\times0.5\times0.5046=0.05500$. 
\item[(ii)] ``hydro + minijet + dijet'' model: we include minijet clusters in addition to (i). The calculated $\vtwo{bg}$ by Eq.~(\ref{eqn13}) is $\vtwo{bg}=\frac{150}{203.8}\times0.05+\frac{5\times10}{203.8}\times0.2\times0.6065+\frac{2\times1.9}{203.8}\times0.5\times0.5046=0.07126$. 
\item[(iii)] ``hydro + minijet + near-side + away-side double-peak'' model: we generate jet-correlated particles by correlation function with double-peak away-side to replicate the experimentally measured reaction-plane averaged dihadron correlation function~\cite{Horner,Jia}. We used the same Gaussian parameters for the correlation peaks as in (i) but $\theta=1$, thus $\mean{\cphi}_{jet}=0.1689$. The calculated $\vtwo{bg}$ by Eq.~(\ref{eqn13}) is $\vtwo{bg}=\frac{150}{203.8}\times0.05+\frac{5\times10}{203.8}\times0.2\times0.6065+\frac{2\times1.9}{203.8}\times0.5\times0.1689=0.06813$. 
\item[(iv)] ``hydro + minijet + near-side + reaction-plane dependent away-side double-peak'' model: we include reaction-plane dependent jet-correlation function similar to preliminary experimental data~\cite{Aoqi}. We have to use Eq.~(\ref{eqn11}) to calculate $\vtwo{bg}$, which gives 
$\vtwo{bg}=\frac{150}{203.5}\times0.05+\frac{5\times10}{203.5}\times0.2\times0.6065+\frac{2\times1.74}{203.5}\times0.1423=0.06910$. 
Note that, in this simulation of reaction-plane dependent jet-correlation signal, the number of jet-correlated particles is not 1.9, but rather 1.74. Also note that, due to the reaction-plane dependency of the jet-correlation signal, the elliptic anisotropy of jet-correlated particles cannot be factorized into the product of the trigger particle elliptic flow and the average angular spread of the jet-correlation signal as in Eq.~\ref{eqn12}, but has to be calculated by Eq.~\ref{eqn6p7}.
\end{itemize}

We list our comparison in Table~\ref{tab}. The fit $\vtwo{{\rm fit}}$ is supposed to be the real background $\vtwo{bg}$. The fit errors are due to statistical fluctuations in the simulation. As can be seen, the calculated $\vtwo{bg}$ reproduces the real background $\vtwo{bg}$ in every case. The $\vtwo{bg}$ values differ from the hydro-background $v_2$ due to contributions from cross-talks between cluster correlation and cluster flow. 
Also shown in Table~\ref{tab} are the two-particle $\vv{2}$ from all pairs in inclusive events. The $\vv{2}$ values differ from $\vtwo{bg}$ due to 
non-flow contributions between particles from the same dijet or the same cluster.

\begin{table}
\caption{Monte Carlo verification of analytical results of elliptic flow parameter to be used in jet-correlation background. Hydro-particle multiplicity, trigger particle multiplicity, jet-correlated near- and away-side multiplicities, number of minijet clusters, and particle multiplicity per minijet cluster are all generated with Poisson distributions, with averages $\Nhy$, $\Nt$, $N_{\ns}$, $N_{\as}$, $\Ncl$, and $\Na$, respectively. The jet-correlation function is given by Eq.~(\ref{eqn16}), with near- and away-side Gaussian width fixed to be $\sigma_{\ns}=0.4$ and $\sigma_{\as}=0.7$, respectively. The minijet cluster Gaussian width is fixed to $\sigma_{\acl}=0.5$. The elliptic flow parameters for hydro-particles, trigger particles, and clusters are $\vtwo{hy}$, $\vtwo{t}$, and $\vtwo{cl}$, respectively, and are fixed over all events without fluctuation. We use $\Nhy=150$, $\Nt=2$, $\Na=5$, $\vtwo{hy}=0.05$, and $\vtwo{cl}=0.20$.}
\label{tab}
\begin{tabular}{llccc}\hline\hline
Case & Parameters\hspace{1in} & $\vv{2}$ & $\vtwo{{\rm fit}}$ & Calculated $\vtwo{bg}$\\\hline
(i) hydro + dijet
& $\Ncl=0$, $\vtwo{t}=0.5$, & & &\\
& $C=0$, $N_{\ns}=0.7$, $N_{\as}=1.2$, & 0.05557(8)	& 0.05505(8)	& 0.05500\\
& $\sigma_{\ns}=0.4$, $\sigma_{\as}=0.7$, $\theta=0$ & & &\\
(ii) hydro + minijet + dijet
& $\Ncl=10$, $\vtwo{t}=0.5$, & & &\\
& $C=0$, $N_{\ns}=0.7$, $N_{\as}=1.2$, & 0.08465(6)	& 0.07115(8)	& 0.07126\\
& $\sigma_{\ns}=0.4$, $\sigma_{\as}=0.7$, $\theta=0$ & & &\\
(iii) hydro + minijet + near-side + 
& $\Ncl=10$, $\vtwo{t}=0.5$, & & &\\
 away-side double-peak
& $C=0$, $N_{\ns}=0.7$, $N_{\as}=1.2$, & 0.08172(6)	& 0.06815(8)	& 0.06813\\
& $\sigma_{\ns}=0.4$, $\sigma_{\as}=0.7$, $\theta=1$ & & &\\
(iv) hydro + minijet + near-side +
& $\Ncl=10$, $\vtwo{t}=0.1$, & & &\\
reaction-plane dependent away-side double-peak & $C=0$, $N_{\ns}=0.7$, $N_{\as}=1.2$, & 0.08279 & 0.06883(35) &	0.06910 \\ 
 + clusters
& $\sigma_{\ns}=0.4$, $\sigma_{\as}=0.7$, $\theta=1$ & & &\\\hline\hline
\end{tabular}
\end{table}

\begin{figure*}[hbt]
\centerline{
\psfig{file=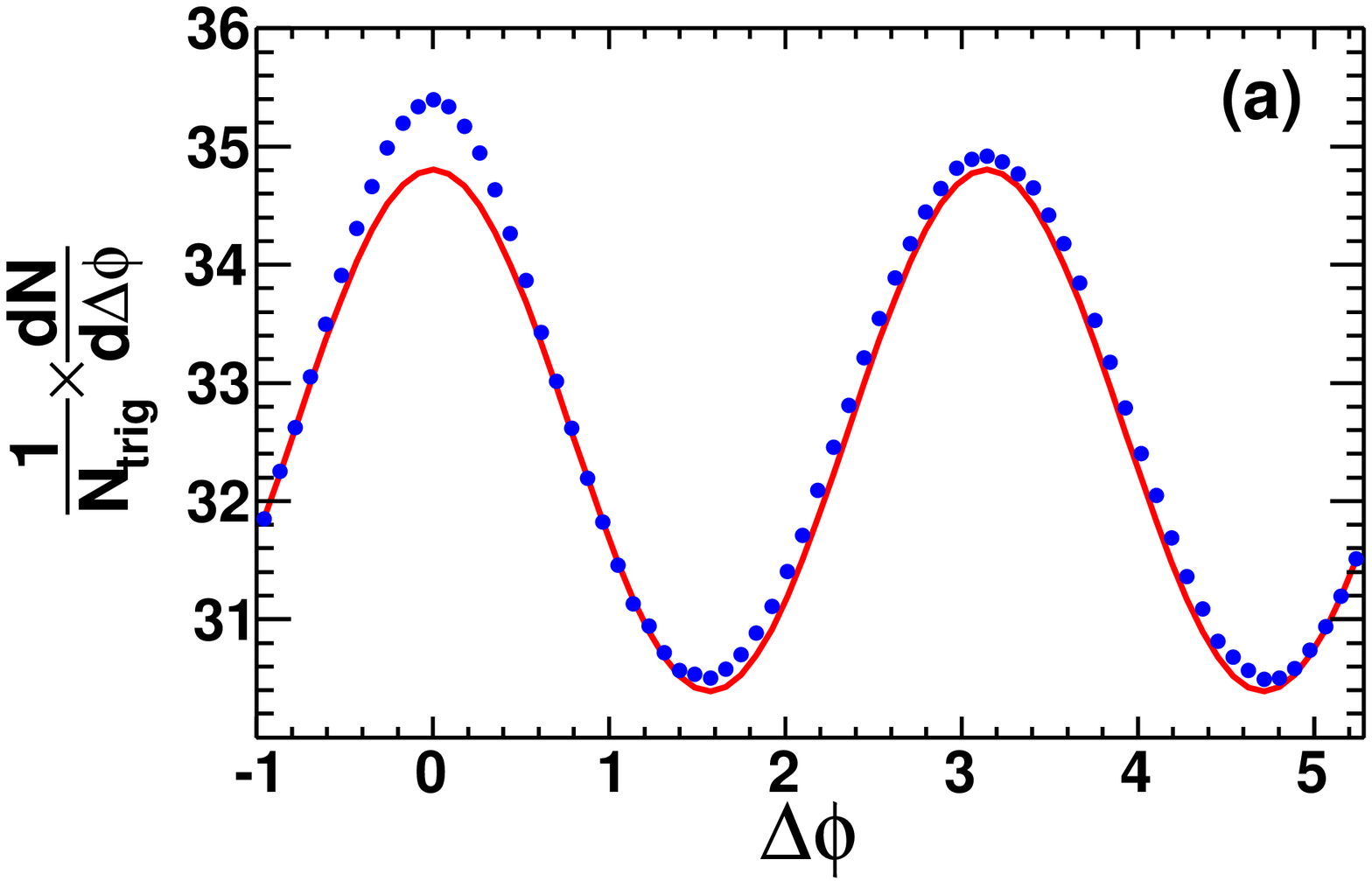,width=0.33\textwidth}
\psfig{file=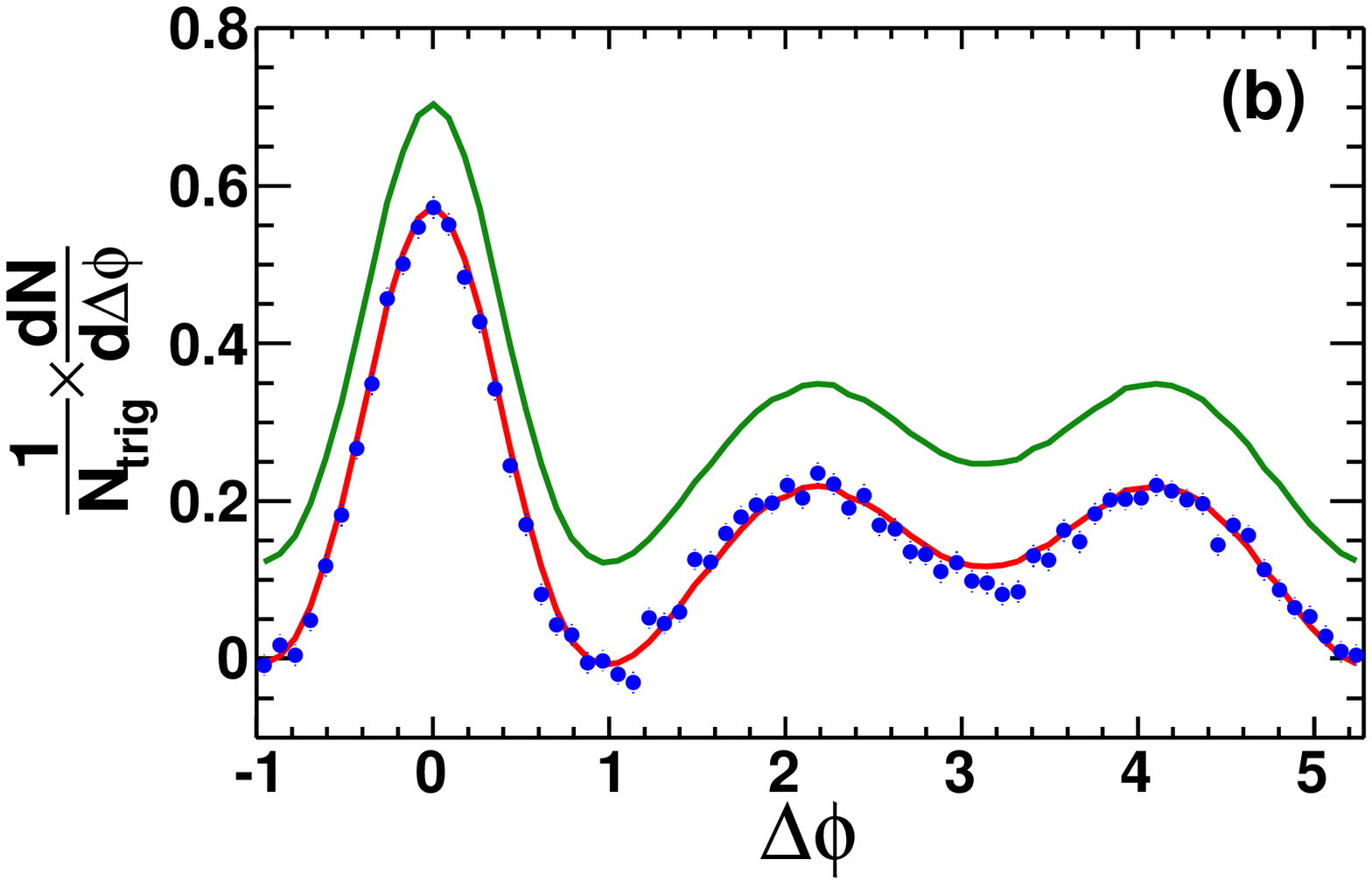,width=0.33\textwidth}
\psfig{file=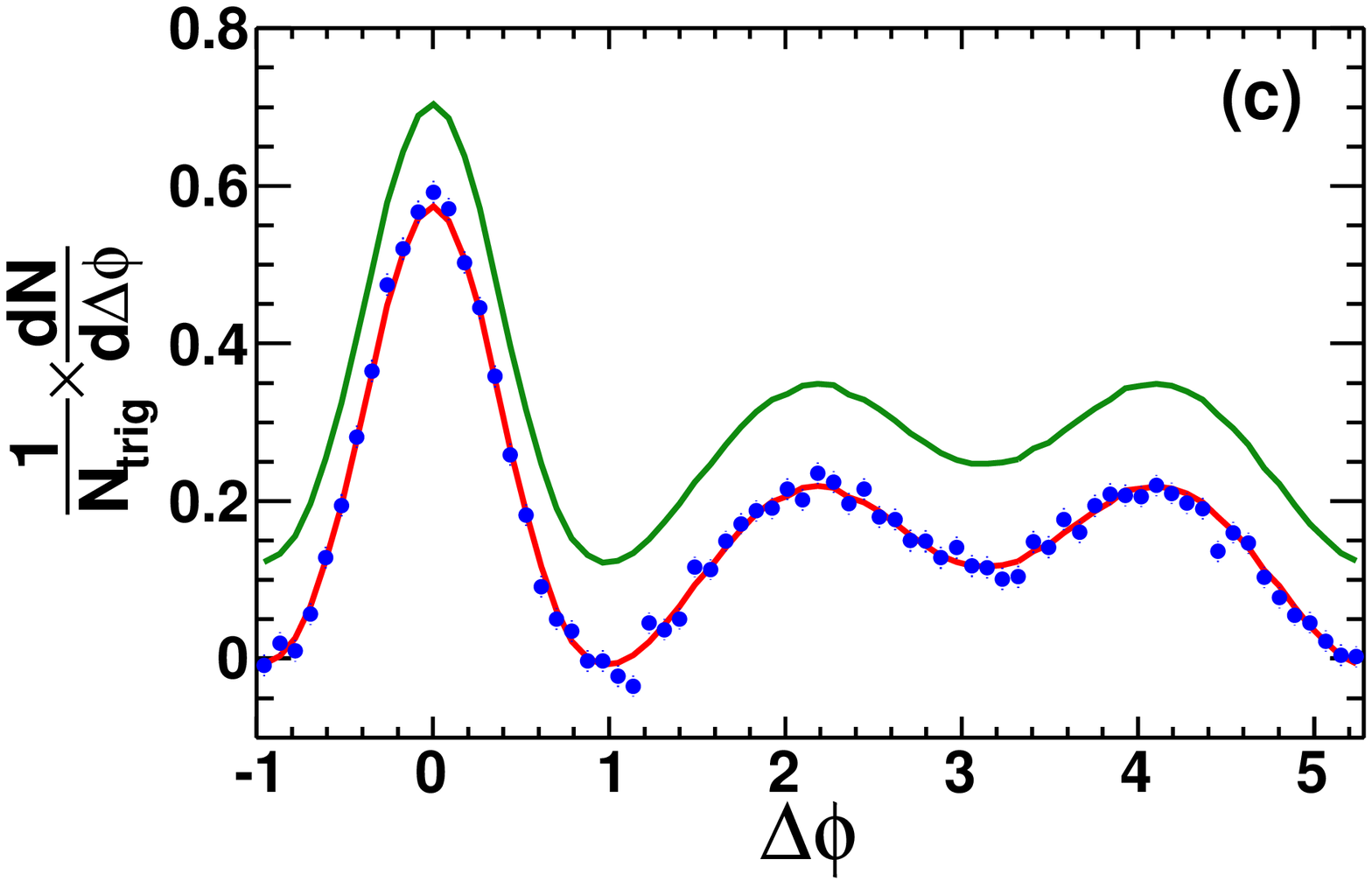,width=0.33\textwidth}
}
\caption{(a) Simulated raw correlation from the ``hydro + minijet + near-side + away-side double-peak'' model (Case iii in Table~\ref{tab}). ZYA1-normalized flow-background using the calculated $\vtwo{bg}$ is shown as the curve. Note the input $v_2$ in the simulation is purposely made much larger than real data to exaggerate non-flow effect. (b) Background-subtracted jet-correlation (data points) compared to the input correlation signal (upper curve). The background uses the calculated $\vtwo{bg}$ and is normalized to signal by ZYA1. The input signal shifted down by a constant is shown in the lower curve. (c) As same as (b) except the subtracted background uses the decomposed $\vv{2D}$.}
\label{fig}
\end{figure*}

Figure~\ref{fig}(a) shows the raw correlation function for case (iii) and flow background using the calculated $\vtwo{bg}$ by Eq.~\ref{eqn11} and normalized by ZYA1. Figure~\ref{fig}(b) shows the ZYA1-background subtracted jet-correlation function, using the calculated $\vtwo{bg}$ for flow background. The background-subtracted jet-correlation is compared to the input signal. As shown in Fig.~\ref{fig}(b), the shapes of the input signal and extracted signal are the same, which is not surprising because the calculated $\vtwo{bg}$ is the correct value to use in flow background subtraction. The roughly constant offset is due to ZYA1-normalization.

\section{How to ``measure'' jet-correlation background (reaction plane) $v_2$}

Two-particle angular correlation is analyzed by STAR and is decomposed into two components~\cite{minijet}: one is the azimuth quadrupole, $\vv{2D}$, that is due to correlations of particles to a common source, the reaction plane; the other is minijet correlation that is due to angular correlation between particles from the same minijet or the same cluster. Of course, any such decomposition is model-dependent; because the functional form for minijet (or cluster) correlation is unknown a priori, one has to make assumptions about its functional form. Modulo this caveat, if cluster correlation and flow correlation is properly decomposed, the azimuth quadrupole should correspond to the first term in r.h.s.~of Eq.~(\ref{eqn14}),
\be
\vv{{\rm 2D}}=\frac{\Nhy}{\Nbg}\vv{2}_{hy}+\sum_{cl}\frac{\Ncl\Na}{\Nbg}\vv{2}_{\acl}.
\ee
This is identical to Eq.~(\ref{eqn11}). That is, the elliptic flow parameter from a {\em proper} 2D quadrupole-minijet decomposition is exactly what is needed for jet-correlation background calculation. 

A 2D quadrupole-minijet decomposition, with an assumption of the minijet correlation structure, has been carried out experimentally by STAR as a function of centrality but including all $\pt$~\cite{minijet}. One may restrict to narrow $\pt$ windows to obtain $\vv{2D}$ as a function of $\pt$, however, statistics can quickly run out with increasing $\pt$ because the 2D decomposition method requires particle pairs.

We perform a decomposition of flow and cluster correlation using our simulation data where the shape of cluster correlation is known from the simulation input. We form two-particle correlation between all particles (untriggered correlation). We fit the two-particle correlation with the sum of cluster correlation and flow to extract $\vv{2D}$ from the simulation data. Figure~\ref{fig}(c) shows the ZYA1-background subtracted jet-correlation function, using the decomposed $\vv{2D}$ for flow background. The background-subtracted jet-correlation is compared to the input signal. The shapes of the input signal and extracted signal are the same, which demonstrates that the decomposed $\vv{2D}$ is close to the real elliptic flow value. Again, the roughly constant offset is due to ZYA1-normalization.

One natural question to ask is why not to decompose jet-correlation and jet-background directly from high-$\pt$ triggered correlation function. One obvious reason is, again, that jet-correlation shape is unknown a priori, thus one cannot simply fit triggered correlation to a given functional form. This is the same caveat mentioned above in particle pair correlation without a special trigger particle, where the minijet shape function has to be assumed in the decomposition of minijet correlation and flow. The situation in jet-like correlation is direr because the main interest of jet-like correlation studies is the investigation of medium modification to jet-correlation structure. Furthermore, even when the functional form of jet-correlation signal is known, as is the case in our simulation, we found that the decomposed jet-correlation signal shape deviates significantly from the input one. This is because the jet-correlation signal is not orthogonal to flow background, but rather entangled, both with near- and away-side peaks, and hence one can get false minimum $\chi^2$ in decomposing the two components with limited statistics. 

\section{Discussions and Summary}

In experimental analysis, $v_2$ values from various methods have been used for jet-correlation background. STAR used the average of the event plane $\vv{EP}$ and the four-particle $\vv{4}$ and used the range between them (or between $\vv{2}$ and $\vv{4}$) as systematic uncertainties~\cite{jetspec,3part}. The event plane $\vv{EP}$ and two-particle $\vv{2}$ contain significant non-flow contributions, while the non-flow contributions are significantly reduced in the four particle $\vv{4}$~\cite{Aihong}. On the other hand, effect of flow fluctuation is positive in $\vv{EP}$ and $\vv{2}$ but is negative in $\vv{4}$. This ensures that the true $v_2$ is smaller than $\vv{EP}$ and $\vv{2}$, and is most likely larger than $\vv{4}$. It is worth to note that the flow parameter to be used in jet-correlation background subtraction should include flow fluctuation effect as in $\vv{2}$, which makes the $\vv{4}$ parameter as the lower limit rather conservative.

The recently measured $\vv{2D}$ magnitudes from STAR are larger than $\vv{4}$ in peripheral and medium central collisions, confirming the validity to use $\vv{4}$ as the lower systematic limit of $v_2$. In central collisions, however, the extracted $\vv{2D}$ is smaller than $\vv{4}$ although the difference is significantly smaller than the difference between $\vv{EP}$ and $\vv{4}$. This would suggest, assuming that the decomposed $\vv{2D}$ reflects the real flow background (i.e., the minijet shapes used in the decomposition is close to reality), that the used $v_2$ values for background calculation in dihadron correlation analysis in STAR would be too large by about $1\sigma$ systematic uncertainty. In three-particle correlation analysis~\cite{3part}, the $\vv{2D}$ was included in the $v_2$ systematic uncertainty assignment. 

STAR has also measured elliptic flow at mid-rapidity in the main Time Projection Chamber (TPC) using event-plane constructed by particles at forward and backward rapidities in the forward TPCs, $\vv{FTPC}$. The obtained $\vv{FTPC}$ is smaller than $\vv{EP}$ using particles from the main TPC only, however, it is still significantly larger than $\vv{4}$. This suggests that some but not all non-flow effects are removed from $\vv{FTPC}$. The remaining non-flow may be dominated by the long range $\Delta\eta$ correlation (ridge) observed in non-peripheral heavy-ion collisions~\cite{jetspec,Joern,PHOBOS}. PHENIX used $\vv{BBC}$ results from the event plane method where the event plane is determined by particles in the Beam-Beam Counter several units of pseudo-rapidity away from particles used in jet-correlation analysis~\cite{PHENIX}. The rapidity gap in the PHENIX measurement of $\vv{BBC}$ is larger than that in the STAR measurement of $\vv{FTPC}$, so non-flow effect should be smaller in the PHENIX measurement. However, it is possible that the $\vv{BBC}$ values used by PHENIX for background calculation can be also too large if non-flow ridge correlation persists to very large pseudo-rapidity gap.

In summary, we have derived an analytical form for jet-correlation flow-background in a cluster approach. We argue that the elliptic flow $v_2$ parameter to be used in jet-correlation background is that from two-particle method excluding non-flow correlation unrelated to the reaction plane, but including cross-terms between cluster correlation and cluster flow. We have verified our result by Monte Carlo simulation for various jet-correlation signal shapes as well as varying other input parameters to the simulation. We demonstrate that the $v_2$ parameter to use in jet-correlation flow background is as same as the $\vv{2D}$ from a proper 2D quadrupole-minijet decomposition of two-particle angular correlation. However, we note that 2D quadrupole-minijet decomposition requires a model for minijet correlation shape, which gives rise to systematic uncertainty on the extracted $\vv{2D}$ which require further studies.

\section*{Acknowledgment}

This work is supported by U.S. Department of Energy under Grant DE-FG02-88ER40412.


\end{document}